\documentclass[final,5p,times,twocolumn,amsmath,sort&compress]{elsarticle}

\usepackage{epsfig}

\usepackage{amssymb}

 \usepackage{graphicx}
\usepackage{dcolumn}
\usepackage{bm}
\usepackage{color}
 \usepackage{lineno}
\usepackage{hyperref}
\usepackage{multirow}
\usepackage{multicol}
\usepackage{amsmath}
\usepackage{textcomp}
\usepackage{tabularx,booktabs,multirow}
\usepackage{siunitx}

\journal{Journal of Magnetic Resonance}

\begin{document}

\begin{frontmatter}



\title{CW and pulsed electrically detected magnetic resonance spectroscopy at 263 GHz/12 T on operating amorphous silicon solar cells}


\author[label1]{W. Akhtar\corref{cor1}}
\cortext[cor1]{ Address : Kekul\`{e}stra{\ss}e 5, 12489 Berlin, Germany ,Phone: +49-30-8062-41353}
\ead{mohammad.akhtar@helmholtz-berlin.de}

 \author[label1]{A. Schnegg\corref{cor1}}
 \ead{alexander.schnegg@helmholtz-berlin.de}
 \author[label2,label3]{S. Veber}
 \author[label4]{C. Meier}
 \author[label1]{M. Fehr}
 \author[label1]{K. Lips}
 
 \address[label1]{Berlin Joint EPR Lab, Institut f{\"u}r Silizium-Photovoltaik, Helmholtz Zentrum Berlin f{\"u}r Materialien und Energie, Germany }
 \address[label2]{Laboratory of Magnetic Resonance, International Tomography Center SB RAS, Russia }
 \address[label3]{Novosibirsk State University, Russia}
 \address[label4]{Berlin Joint EPR Lab, Fachbereich Physik, Freie Universit{\"a}t Berlin, Germany}
 
\begin{abstract}
Here we describe a new high frequency/high field continuous wave and pulsed electrically detected magnetic resonance (CW EDMR and pEDMR) setup, 
operating at 263 GHz and resonance fields between 0 and 12 T. Spin dependent transport in illuminated hydrogenated amorphous silicon p-i-n solar 
cells at 5 K and 90 K was studied by in operando 263 GHz CW and pEDMR alongside with complementary X-band CW EDMR. Benefiting from the superior resolution at 263 GHz, 
we were able to better resolve EDMR signals originating from spin dependent hopping and recombination processes. 5 K EDMR spectra were found to be dominated by conduction 
and valence band tale states involved in spin dependent hopping, with additional contributions from triplet exciton states. 90 K EDMR spectra could be assigned to spin pair 
recombination involving conduction band tail states and dangling bonds as dominating spin dependent transport process, 
with additional contributions from valence band tail and triplet exciton states.

\end{abstract}


\end{frontmatter}


\section{INTRODUCTION }

Electrically detected magnetic resonance (EDMR) spectroscopy is a highly sensitive spectroscopic technique that probes the involvement of paramagnetic states
in charge transport and loss mechanisms, while paramagnetic background signals are eliminated~\cite{DJL,KSM,BL}. EDMR is up to 8 $\sim$~10 orders of magnitude 
more sensitive than conventional electron paramagnetic resonance (EPR) spectroscopy~\cite{DRM} and thus allows for investigating a small number of spins in low dimensional 
material systems and devices~\cite{DRM,XMY,UMO,LPM,PBM}. In addition, EDMR is regarded as one of the key technologies to realize quantum computing algorithms based on 
coherent manipulation and readout of electron spins in semiconductor samples~\cite{SBH,MAA,LVM,LLM}. Due to its high selectivity and sensitivity, EDMR is 
ideally suited for the assignment and structural characterization of paramagnetic states determining charge transport and loss mechanisms in organic and
Si solar cells~\cite{LF,SBB,LKF,MKV,KLF,GBS,MBT,KLCB}.
Recently, it was shown that the application range of EDMR as compared to continuous wave (CW EDMR) may be further boosted by pulsed (pEDMR) detection schemes~\cite{SBH,LVM, BK,BKL,BLB,HHG,HHS,HDH}, which greatly increased the selectivity to different spin-dependent transport mechanisms and spin 
coupling parameters as well as the spectral resolution~\cite{SBF,JBS,JBT,HLM,FFCB,CBKM,MFJB,ARJ,SKSB,MTAF}. 
Up to now most CW and pEDMR studies employed conventional X-band (9.4 GHz/350 mT) spectrometers. However, the spectral resolution at this field is 
often not sufficient to resolve g-tensor anisotropies and the resonances of different overlapping paramagnetic states, having close proximity in g-values. 
This restriction may be lifted by high frequency/high field EDMR, which is capable of resolving small g-tensor anisotropies and overlapping EDMR resonances 
with slightly different g-values when the anisotropic Zeeman interaction, ($\Delta$g/g$_\text{iso}$)B$_0$ exceeds the field independent inhomogeneous broadening, 
$\Delta$B, i.e. ($\Delta$g/g$_\text{iso}$)B$_0$ $> \Delta$B, where $\Delta$g is the g-value difference, g$_\text{iso}$ is the mean g-value and B$_0$ is the external magnetic field.  

A very first report on high frequency/high field CW EDMR appeared as early as 1978 in which Honig $et$ $al.$ studied phosphorus doped crystalline 
Si (c-Si:P) at 196 GHz/7 T and a temperature of 1.4 K~\cite{HM}. This work was extended in a later 240 GHz/8.6 T EDMR study on
c-Si:P at T = 2.8 K~\cite{MMSB,MTM}, where it was shown that the dominating spin-dependent transport process at very high fields and 
low temperatures is a direct capture of conduction band electrons by phosphorus states, while spin-pair formation dominates the low-field EDMR spectra in 
this system~\cite{SBH}. This important finding demonstrates that high frequency/high field EDMR increases not only the spectral resolution, but 
also the selectivity to particular transport processes. 

\begin{figure*}[t]
\begin{center}
\includegraphics[width=14cm]{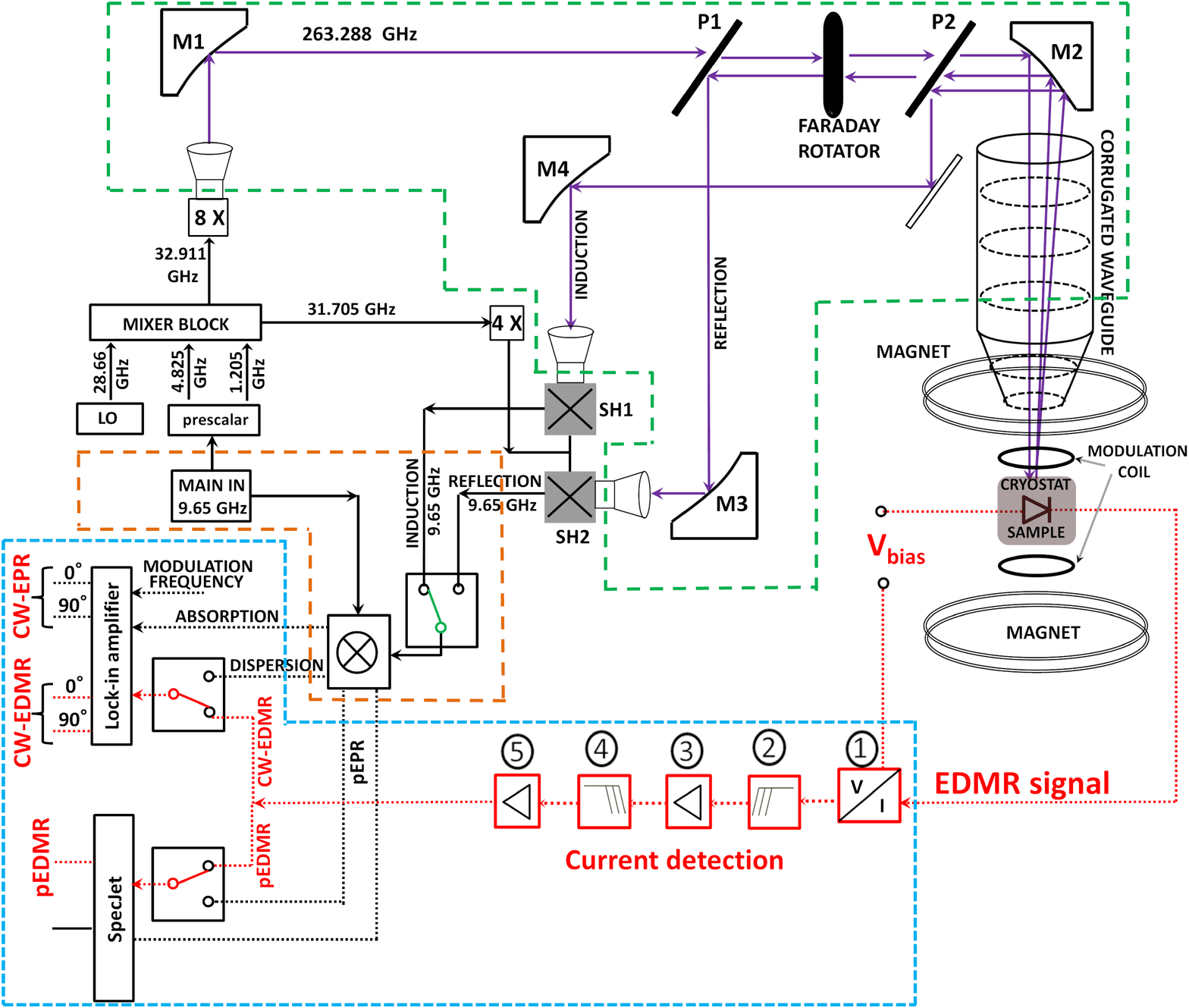}
\caption {(color online). Scheme of the CW/pulsed 263 GHz EPR/EDMR spectrometer. The basic spectrometer unit is an X-band mw bridge 
(orange dashed block) which determines mw frequency, phase and time structure. A mixer block converts X-band to Q-band frequencies, 
which are further multiplied to achieve the final 263 GHz frequency. 263 GHz radiation is than emitted into the quasi-optical bridge
(green dashed block) by horn antennas, where it propagates as a Gaussian beam (solid purple line). Off-axis elliptical mirrors (M1 and M2) 
direct the 263 GHz Gaussian beam into the corrugated waveguide for sample excitation inside the magnet. Wire-grid polarizers (P1, P2) and a 
Faraday rotator separate induction and reflection mode EPR signals, which are down converted by sub-harmonic mixers (SH1 and SH2) to X-band frequencies 
and further processed by the detection electronics (blue dotted block). For EDMR detection the current through the solar cell 
(represented as a diode inside the magnet) is detected by a (1) trans-impedance amplifier, (2) a high-pass filter, 
(3) an amplifier, (4) a low-pass filter and (5) an additional amplifier. The final electrical signal is than fed into the detection electronics. }  
\label{fig:one}
\end{center}
\end{figure*}

A further W-band (94 GHz/3.4 T) EDMR study on a donor-doped metal-oxide-semiconductor device employing a single mode cavity at 5 K was reported 
by Lo $et$ $al.$~\cite{LLM, LLG}. In this study sample dimensions were of sub-millimeter range to fit a W-band single mode resonator. 
Due to the superior resolution of high frequency/high field EDMR, the donor and the two dimensional electron gas (2-DEG) EDMR spectra, which overlap 
at X-band frequencies, could be separated. 
Very recently Meier $et$ $al.$ conducted a multi-frequency (X-band, Q-band and W-band) pEDMR study on micro-crystalline ({\textmu}c-Si:H) p-i-n solar cells
with a-Si:H n-layer~\cite{MBT}. Thereby, the g-values and line shapes of a-Si:H conduction band tail states  and localized {\textmu}c-Si:H conduction band (CE) states 
could be determined with increased resolution. 
In order to further push the capabilities of very high frequency/high field EDMR, with respect to the accessible resonance frequencies/fields and
temperatures, within this work we developed a CW EDMR/pEDMR set-up based on a commercial Bruker E-780 1016 EPR spectrometer working 
at 263 GHz/12 T. Herein, we give a detailed description of this instrument and demonstrate first 263 GHz CW EDMR and pEMDR experiments on
operating a-Si:H p-i-n solar cells.

\section{EDMR SPECTROMETER $@$ 263 GHz}

\begin{figure}[t]
\begin{center}
\includegraphics[width=8.5cm]{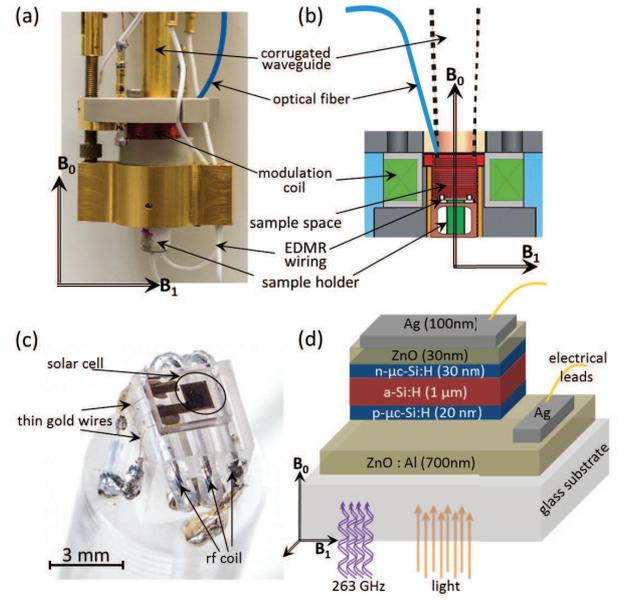}
\caption {(color online). (a) Photograph and (b) cross section of the 263 GHz EDMR probe. (c) Photograph of the EDMR sample holder with mounted 
a-Si:H p-i-n solar cell. The radio frequency (rf) coils shown in the figure can be used for rf excitation of the sample. (d) Layer structure of the a-Si:H p-i-n solar cell.  
B$_0$ and B$_1$ indicate the constant external magnetic field and the magnetic component of the mw field, respectively. }  
\label{fig:two}
\end{center}
\end{figure}

Fig.~\ref{fig:one} shows the layout of the excitation and detection scheme of the 263 GHz spectrometer including the newly implemented 
CW EDMR/pEDMR extension. The depicted EDMR setup is based on a commercial Bruker E-780 1016, 263 GHz/12 T EPR spectrometer, which can be 
operated in CW and pulsed EPR and electron-nuclear double resonance (ENDOR) modes. Microwave (mw) generation and manipulation is done in a heterodyne 
microwave bridge. The corresponding resonance fields are generated in a cryogen-free superconducting magnet, which consists of a main and a sweep coil. 
The main solenoid coil can be swept between 0 and 12 T, while the sweep coil generates additional fields of $\pm$ 0.15 T with high accuracy. The latter can 
be driven in a special linearize mode to compensate for field nonlinearities.
EDMR samples are placed in a non-resonant EPR/EDMR probe which is depicted in Fig.~\ref{fig:two}, with a photograph of the probe in Fig.~\ref{fig:two}a. 
Fig.~\ref{fig:two}b sketches the cross section of the probe showing the corrugated waveguide tapered down to the sample space, the position of the laser fiber 
for light excitation of the sample, the modulation coil (maximum modulation amplitude 50 G $@$ 100 kHz) and the sample position with the electrical contacts for 
EDMR detection. Inside the probe, sample temperature can be controlled between 4.5 K to 300 K using an Oxford ITC 503 temperature controller.

Fig.~\ref{fig:two}c shows the p-i-n solar cell sample of dimension $1\times1$ mm (active solar cell area) that is attached to the sample holder. Electrical connections to the solar cell 
are realized by thin gold wires fixed using silver paste on the contact pads and finally soldered to the coaxial line, which is used to apply a constant 
voltage and to monitor the photocurrent from the sample. The layer structure of the p-i-n solar cell is depicted in Fig.~\ref{fig:two}d. It consists of a 
1 {\textmu}m thick intrinsic a-Si:H layer, which is sandwiched between thin p-type (20 nm) and n-type (30 nm) {\textmu}c-Si:H layers and Al doped ZnO layer, that act 
as transparent conducting oxide contacts. We have used  {\textmu}c-Si:H  as contact layers to distinguish signals from the p and n doped layers from that
of the intrinsic a-Si:H layer as the two materials have different EDMR resonances~\cite{LF,SBB,JBS}.
Photo excitation of the solar cell is achieved by laser sources or the light from a 250-watt KL 2500 LCD halogen cold light source through a 1.22 mm optical 
fiber. The optical illumination and mw excitation of the solar cell is done through the glass substrate as depicted in the Fig.~\ref{fig:two}d. 
The details of the current detection setup are shown in Fig.~\ref{fig:one}. Here a steady-state photocurrent is established under a constant voltage bias 
(V$_\text{bias}$) and illumination. The output of the trans-impedance amplifier (I-V convertor) converts the photocurrent into a voltage which goes through the 
filtering and amplification circuit (shown in the Fig.~\ref{fig:one} in red blocks) to maximize the signal-to-noise ratio.
This set-up allows for both CW EDMR as well as pEDMR detection. For CW EDMR detection, the current detection output is connected to either the absorption or
dispersion channel of the Lock-In detector. Since the output of the Lock-In provides 4 channels (0$^{\circ}$ and 90$^{\circ}$ of the EPR absorption and 
dispersion signals, respectively), EPR and EDMR signals can be measured simultaneously. For pEDMR we directly connect the output of the current detection 
to one of the channels of the Bruker SpecJet digitizer (bandwidth 500 MHz). The maximum mw power available in CW as well as in pulse mode is 15 mW. The current 
transient after the mw pulse is boxcar integrated and  recorded as a function of magnetic field to yield the pEDMR spectrum~\cite{BL}.

The current set-up allows for pulsed and CW EDMR with magnetic field and mw amplitude modulation as well as electrically detected electron electron double resonance (EDELDOR)~\cite{HHS}, electrically 
detected electron nuclear double resonance (EDENDOR)~\cite{HDH} and electrically detected double electron electron resonance (EDDEER)~\cite{SKSB}.
These experiments carried out without the employment of a resonator holds great advantage as fully processed samples of several mm size may be mounted, 
temperature induced instabilities are less pressing as compared to single mode resonator set-ups and in addition it provides more freedom to add additional 
electrical connections, rf coils or illumination facilities.
However, omitting a resonator significantly reduces the mw field at the sample position. For the solar cell used in this study, mw excitation has to be done through the glass 
substrate (see Fig.~\ref{fig:two}d) which further leads to nearly 10 dB loss in mw power. Even with the above mentioned losses, the minimum $\pi$/2 pulse length achievable in the current EDMR set-up 
(15 mW at the entrance of the corrugated waveguide) is $\approx$ 1 {\textmu}s, which corresponds 
to an excitation bandwidth of $\approx$ 1 MHz. This puts a restriction to perform pEDMR on samples with short relaxation times as well as on the detection of electron spin echo
envelop modulation, ESEEM 
(nuclear Larmor frequency at 9.4 T for $^{15}$N $\approx$ 41 MHz, $^{29}$Si $\approx$ 80 MHz,  $^{31}$P $\approx$ 162 MHz and 
$^{1}$H $\approx$ 400 MHz).
In future, changing the sample design such that the mw excitation can be realised from the front side, we will be able to avoid the loss due to glass substrate. With such
a design shorter $\pi$/2 pulses 
will be possible that will increase the application range of pEDMR experiments.

The possibility to measure EPR and EDMR simultaneously in the current set-up greatly improves the accuracy of field calibrations, since for the precise 
determination of g-values reference EPR measurements with field standards (in our case Mn$^{2+}$ in MgO matrix~\cite{BRG}  or N$@$C$_{60}$~\cite{PRV}) are
required. These calibrations only yield optimum precision if the reference EPR sample can be placed exactly at the same position as the EPR/EDMR sample
under study. Mounting and demounting of the reference sample may fail to reproduce the sample position. In order to circumvent this problem we attached a 
small amount of N$@$C$_{60}$ underneath the solar cell and recorded its EPR alongside with the EDMR.

 \section{EXPERIMENTAL RESULTS AND DISCUSSION}
 
\begin{figure}[t]
\begin{center}
\includegraphics[width=8.50cm]{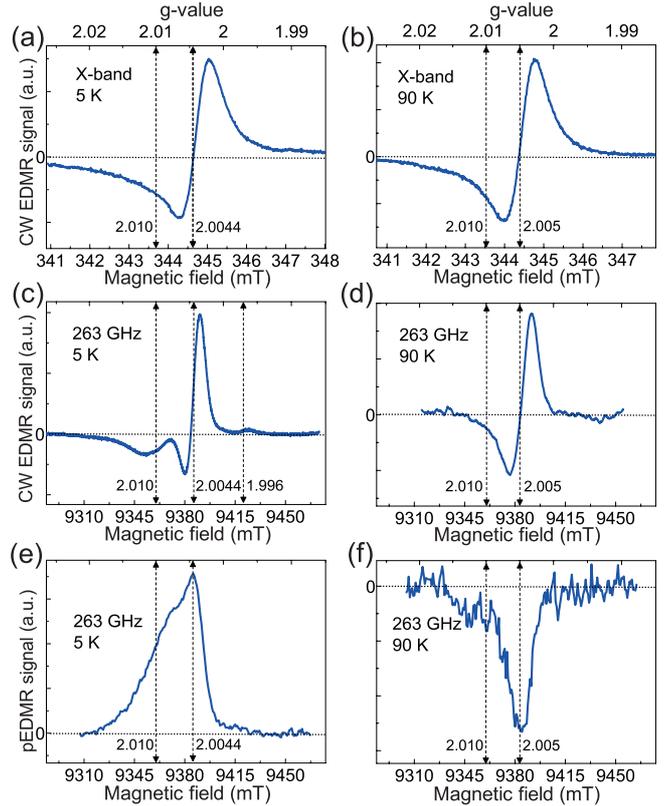}
\caption {(color online). X-band CW EDMR (first row), 263 GHz CW EDMR (second row) and 263 GHz pEDMR (third row) spectra obtained at 5 K (left column) and 
90 K (right column), respectively on a-Si:H p-i-n solar cell. The mw frequency for (a) and (b) is 9.6688 and 9.6644 GHz, respectively, and 263.32 GHz for (c) to (f). 
CW spectra at 5 K and 90 K were taken with the same Lock-in phase. Spectra are plotted vs. magnetic field (bottom axis) and g-values (top axis). pEDMR spectra were 
corrected for microwave induced current changes by the subtraction of off-resonance current transients. Linear baseline correction was applied to 263 GHz CW EDMR spectra. 
Main peaks in the spectra are marked by vertical double sided arrows.} 
\label{fig:three}
\end{center}
\end{figure}

Fig.~\ref{fig:three} depicts X-band CW and 263 GHz CW and pEDMR spectra obtained on illuminated a-Si:H solar cells under forward bias (V = +1 V) at 5 K and 90 K, 
respectively. For better comparability, all spectra are plotted against the same g-value span (upper x-axes) and the corresponding magnetic field values 
(bottom x-axes). In the following, we will first qualitatively describe the characteristics and differences of the EDMR spectra shown in Fig.~\ref{fig:three}.
Based on these findings we will than conclude on the underlying spin dependent transport processes and finally test these assignments by spectral simulations. 

X-band CW EDMR spectra measured at 5 K (Fig.~\ref{fig:three}a) and 90 K (Fig.~\ref{fig:three}b) are very similar. Both spectra are dominated by asymmetric 
lines centered at g = 2.0044 and g = 2.005, respectively, which extend towards larger g-values around g = 2.01. 
In addition, we find a very broad resonance (width $\approx$ 20 mT) centered around g = 2.008, 
which clearly exceeds the spectral window depicted in Fig.~\ref{fig:three}a and ~\ref{fig:three}b. This contribution is more pronounced in the 5 K 
spectrum as compared to 90 K. 
Differences between 5 K and 90 K CW EDMR spectra obtained on the same solar cells under similar operating conditions strongly increase upon 
increasing the resonance frequency from X-band to 263 GHz. The 263 GHz CW EDMR spectrum measured at 5 K (Fig.~\ref{fig:three}c), now exhibits three well resolved 
contributions around g = 2.01 and g = 2.0044 and a smaller signal at g = 1.996. 
Plotting 5 K X-band and 263 GHz 
CW EDMR spectra on the same g-scale reveals a pronounced increase of the spectral resolution, despite of the fact that the individual spectral 
contributions significantly broaden. On the contrary, 90 K CW X-band and 263 GHz spectra (see Fig.~\ref{fig:three}b and ~\ref{fig:three}d) have very similar 
shapes. Their main difference is an increase of the absolute spectral width at 263 GHz.

Observed temperature induced differences between 5 K and 90 K 263 GHz CW EDMR spectra are a clear indication that the spectra are dominated by different 
paramagnetic states and hence different spin dependent transport processes, despite of their similarity at X-band frequencies. Additional information on the 
prevailing spin dependent transport process may be obtained from complementary pEDMR measurements. 5 K and 90 K 263 GHz pEDMR spectra are depicted in 
Fig.~\ref{fig:three}e and ~\ref{fig:three}f, respectively. 263 GHz pEDMR and CW EDMR spectra exhibit identical spectral contributions. However, in addition to 
temperature induced spectral changes 5 K and 90 K pEDMR spectra exhibit different signs. At 5 K, 263 GHz pEDMR induces a current enhancement (positive signal), 
whereas at 90 K pEDMR current quenching (negative signal) is observed.

We will now identify the observed resonances with paramagnetic states and assign them to spin dependent transport pathways. 
Intrinsic a-Si:H is a disordered semiconductor with localized states below the conduction and above the valence band, which are referred to as conduction 
band tail (cbt) and valence band tail (vbt) states. Under illumination cbt and vbt become paramagnetic~\cite{KBS,SBW,DSS,UYI,UMYI}. 
For cbt states in a-Si:H, multi-frequency pEDMR up to W-band frequencies revealed a purely isotropic g-value (g = 2.0047(2))~\cite{MBT}. 
For the vbt counterpart, the situation is less clear. This state is typically associated with a resonance at g = 2.011~\cite{KBS,SBW,DSS}.
However, Q-band EPR on illuminated intrinsic a-Si:H powders revealed pronounced g-tensor anisotropy for vbt states 
(g$_{xx}$ = 2.019, g$_{yy}$ = 2.012 and g$_{zz}$ = 2.005)~\cite{UMYI}. Furthermore, it was concluded that vbt g-values depend on the 
p-type doping level and the sample temperature~\cite{DSB}. However, till date latter assignments have not been confirmed by high frequency EPR or EDMR. 

At liquid He temperatures spin dependent hopping transport along cbt and vbt gives rise to EDMR induced current 
enhancement spectra~\cite{LF,MBMS,KLSW,MKV}.
We therefore assign the 5 K EDMR enhancing signals plotted in Fig.~\ref{fig:three}a, ~\ref{fig:three}c and ~\ref{fig:three}e, to two independent spin 
dependent hopping processes along cbt (g = 2.0044) and vbt (g = 2.011) states, respectively. The small signal at g = 1.996 observed 
in 5 K 263 GHz CW EDMR spectra (Fig.~\ref{fig:three}c) can be 
assigned to a similar hopping process along CE states in the n-doped {\textmu}c-Si:H layer~\cite{KLF,JBS,MFC,MMT}. An additional very broad 
resonance 
(20 mT) centered around g = 2.008 has been observed by optically detected magnetic resonance~\cite{HLM,MORI} and CW EDMR~\cite{MBMS,KLSW}
on undoped a-Si:H layer and was associated to strongly coupled triplet electron-hole (e-h) pairs. 
This triplet exciton signal may account for the broad background observed in the X-band CW EDMR spectrum (Fig.~\ref{fig:three}a). 

Intrinsic a-Si:H exhibits further pronounced EPR and EDMR resonances, which have been associated to three fold
coordinated Si atoms, so called dangling bonds (dbs)~\cite{DSS,UMYI,SDKB,MFBR}. At temperatures, between 70 K and room temperature, tunneling of trapped cbt 
electrons into neutral dbs leads to doubly occupied db (db$^{-}$)
states which finally recombine with holes~\cite{LF,MBMS,KLSW,MKV}. 
The same spin dependent recombination process gives 
rise to the 90 K EDMR spectra depicted in Fig.~\ref{fig:three}b,~\ref{fig:three}d and ~\ref{fig:three}f. 
The peak of the main resonance position at g = 2.005, is a superposition of the db (g = 2.0055) and cbt (g = 2.0044) resonances. 
In pEDMR, this process leads to a current quenching signal (see Fig.~\ref{fig:three}f). 
It should be noted here that the sign change between 5 K and 90 K is not evident in the CW EDMR spectra mainly because the phase of the CW EDMR 
spectra has complex dependence on experimental conditions such as temperature, microwave power and modulation frequency~\cite{SYL}. 
90 K EDMR spectra still show 
contributions from vbt resonances. This is most obvious in the 90 K X-band CW EDMR spectrum depicted in Fig.~\ref{fig:three}b, but can be also found in the 90 K 263
GHz pEDMR current quenching spectrum (Fig.~\ref{fig:three}f). 
Similar signals were previously assigned to a recombination process, where the final electron-hole recombination step occurs between a doubly occupied db$^{-}$ and a hole trapped in a vbt state. This step may be enhanced by spin-dependent diffusion of holes via vbt states towards the db$^{-}$~\cite{DSS}.

To further test these assignments, we performed numerical simulations of the 
experimental EDMR spectra using Easyspin, a
Matlab EPR simulation toolbox~\cite{SSS}. In the simulations, the individual EDMR resonances were modeled by isolated spin centers, for which only the 
respective Zeeman interaction was explicitly taken into account in the Spin Hamiltonian. Any spin-spin interaction between different centers was neglected. Line width parameters for 
S = 1/2 centers (cbt, vbt and db) were modeled using site-to-site g-value variations (g-strain) in combination with field independent Gaussian or 
Lorentzian lines. The latter combination leads to Voigtian line shapes at X-band frequency, with increasing Gaussian contribution at 263 GHz. 
For triplets (S = 1), a field independent dipolar broadened Gaussian line shape was assumed. In cases where independent processes contribute to the
EDMR spectrum, e.g. for spin dependent hopping among vbt and cbt states, the weights of the resonances were determined individually. However, for spectra 
in which two resonances contribute to the same process, e.g. spin dependent recombination from cbt to dbs, the same weights were used for both spin pair 
partners.

  \begin{figure}[t]
\begin{center}
\includegraphics[width=8.5cm]{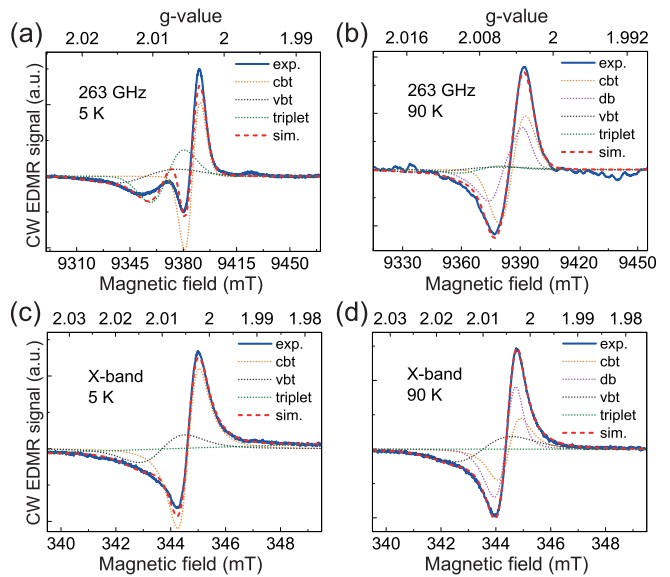}
\caption {(color online). Experimental (solid blue) and simulated (red dashed) CW EDMR spectra. 5K (a) 263 GHz and (c) X-band EDMR spectra were simulated assuming a 
superposition of isotropic cbt (g = 2.0044, g-strain = 0.0025, field independent Lorentzian line width = 1 mT), vbt (g = 2.011, g-strain = 0.01, field independent 
Lorentzian line width = 1.5 mT) and triplet (g = 2.008, field independent Gaussian line width = 26 mT) signals. 90 K (b) 263 GHz and (d) X-band EDMR spectra were 
simulated assuming identical g-values and line shape parameters for cbt, vbt and triplet states and an additional db (g$_{xx}$ = 2.0079, g$_{yy}$ = 2.0061 and 
g$_{zz}$ = 2.0034, g$_{xx}$-strain = 0.0054, g$_{yy}$-strain = 0.0022, g$_{zz}$-strain = 0.0018, field independent Voigtian line of 0.13 mT Gaussian and 0.43 mT Lorentzian 
line width) contribution. Dotted orange, black, magenta and green lines represent 
contributions from cbt, vbt, db and triplet states, 
respectively.}
\label{fig:four}
\end{center}
\end{figure}

Since 5 K and 90 K EDMR spectra are resulting from different spin dependent processes, different models were employed to simulate low and high 
temperature EDMR spectra. Based on the above assignments, we started to simulate the 5 K 263 GHz CW 
EDMR spectrum as a superposition of isotropic cbt (g = 2.0044) and vbt (g = 2.011) states and an additional contribution from triplet excitons (g = 2.008). 
To adjust the line 
intensities and widths of the simulation to the experimental 5K 263 GHz CW EDMR spectrum, a fitting routine was employed in which the line widths and weights were left open. 
Fig.~\ref{fig:four} depicts simulated and experimental EDMR spectra alongside with the individual contributions of different spin centers. Best fits to the 5K 263 GHz 
CW EDMR spectrum (Fig.~\ref{fig:four}a) were achieved using a field independent Lorentzian line with full-width-half-maxima of 1 mT (1.5 mT), field dependent 
g-strain of 0.0025 (0.01) for cbt 
(vbt) lines and a 26 mT broad Gaussian line for the triplet exciton signal. 
Using the same g-values and line width parameters, but different weights for the single paramagnetic states, we were also able to reproduce the main features 
of 5 K X-band CW EDMR spectrum (Fig.~\ref{fig:four}c). However, in both cases we find deviations between simulated and experimental EDMR spectra. 
These differences are most pronounced at the low field edges of the spectra in the range of the vbt resonances. To test whether these discrepancies may originate 
from an anisotropic vbt g-tensor, we performed additional spectral fits to the 263 GHz CW EDMR spectrum assuming isotropic cbt and triplet signals, but anisotropic 
vbt g-values. In the fits all three main components of the vbt g-tensor and the isotropic cbt g-values were left open. A much better agreement between experimental and 
calculated spectra, as compared to the purely isotropic assumption, were obtained for a cbt g-value = 2.00445(6) and an anisotropic vbt g-tensor 
with g$_{xx}$ = 2.015(1), g$_{yy}$ = 2.009(1) and g$_{zz}$ = 2.0058(2) (not shown). 

The observed vbt g-tensor anisotropy is in accordance with previous Q-band EPR experiments on illuminated a-Si:H powders [47]. However, vbt g-values, extracted from 
spectral fits to the 263 GHz CW EDMR spectra, significantly deviate from those extracted by EPR~\cite{UMYI}. Notably, the largest canonical g-value, g$_{xx}$ = 2.015 
deviates from 
g$_{xx}$ = 2.019, the value obtained in Ref.~\cite{UMYI}. Despite the fact that our 263 GHz EDMR approach spreads g-values over a much wider spectral window and 
thereby potentially provides more accurate g-values, different reasons may be responsible for the observed discrepancy. It may be due to different experimental 
conditions and detection 
techniques or the presence of additional yet unassigned paramagnetic states involved in other spin dependent transport processes. In particular, the latter point cannot 
be unambiguously discriminated from a vbt g-tensor anisotropy based on the experimental data presented herein. 

Finally, we simulated the CW EDMR spectra obtained at 90 K assuming spin dependent recombination between cbt and db states as dominating process, with additional 
contributions from vbt states and triplet excitons. g-values, g-strains and field independent line width parameters of cbt, vbt and triplet states were chosen identical 
to those extracted from spectral fits (based on isotropic g-values of cbt and vbt states) to the 5 K 263 GHz EDMR spectrum (see above). The db g-tensor
(g$_{xx}$ = 2.0079, g$_{yy}$ = 2.0061 and g$_{zz}$ = 2.0034) and strains were taken from Ref.~\cite{MFBR}. To simulate the experimental 90 K 263 GHz and X-band CW EDMR 
spectrum, a fitting routine was employed in which the weights of the individual lines were left open, with the restrictions that cbt and db weights are assumed to be the same. 
Fig.~\ref{fig:four}b and ~\ref{fig:four}d depict experimental 263 GHz and X-band 90 K CW EDMR spectra, respectively alongside with the contributions of the individual 
spin centers. Good match between simulated and experimental EDMR spectra confirm the assignment of paramagnetic states contributing to the spectrum.

\section{CONCLUSION}
In conclusion, we have developed a 263 GHz/12 T CW/pEDMR set-up which allows for in-situ measurements on fully processed electronic devices under light and 
applied bias voltage. With this set-up, EDMR measurements can be realized from liquid He to room temperatures. Precise magnetic field calibration at the sample 
position can be obtained by simultaneous EPR measurements of reference samples attached to EDMR samples.

Based on these achievements in EDMR instrumentation we performed first 263 GHz CW/pEDMR on a fully processed a-Si:H p-i-n solar cell. In comparison with X-band CW EDMR, 
these measurements allowed for an assignment of the dominating paramagnetic centers and spin dependent transport processes at 5 K and 90 K. 
At 5 K EDMR signals could be identified with hopping transport along cbt and vbt states and an additional contribution from triplet states. At 90 K spin dependent 
recombination from cbt into dbs was found to dominate, superimposed by additional vbt and triplet signals. 
Spectral simulations based on these assumptions and line shape parameters obtained from literature reproduced the experimental X-band and 263 GHz spectra differently well. 
90 K X-band and 263 GHz data could be simulated very well. However, we found discrepancies between experimental and simulated 5 K EDMR spectra assuming 
contributions from cbt, vbt and triplet states. 
First tentative simulations revealed that this discrepancy may either originate from a pronounced anisotropy of the vbt g-tensor or the contribution of additional 
yet unassigned paramagnetic states. A clear cut assignment of the involved centers requires additional pulsed multi-frequency EDMR experiments, which would allow for a 
discrimination between S = 1/2 and S = 1 spins~\cite{JBT,GA}. Such studies are currently on the way in our lab. Additional evidence may be gained from complementary advanced 
density functional theory calculations, which became available only recently and were employed for the g-tensor calculations of cbt and db in a-Si:H~\cite{GBS,MFBR,ARJ}. In 
the present case, calculations of vbt g-values could provide important additional information on whether vbt states indeed exhibit anisotropic g-values.
The presented results demonstrate that multi-frequency EDMR, including very high frequencies up to 263 GHz, can help to resolve overlapping EDMR resonances and to assign 
transport processes involving the same or different paramagnetic states. 
Combing the high versatility of pEDMR with respect to mw excitation schemes with the ultimate sensitivity of CW EDMR in a high frequency/high field EDMR experiment further 
boosts the potential of this approach. Due to its unmatched detection sensitivity, such studies are not limited to cryogenic temperatures but may in the future also be 
performed at room temperature, which will yield further highly desired pieces of information on spin dependent transport processes in operating electronic devices.

\section{ACKNOWLEDGEMENT}
This work was supported by the German Research Foundation within SPP 1601. We thank F. Finger and O. Astakhov (Forschungszentrum J{\"u}lich) for supplying the 
solar cells which were prepared through the BMBF funded network project EPR-Solar (03SF0328). We are also grateful to J. Behrends (FU Berlin $/$ HZB) for helpful discussions.

\hspace{20 mm}

\end{document}